\documentclass[%
 reprint,
 showpacs,preprintnumbers,
 amsmath,amssymb,
 aps,
]{revtex4-2}

\usepackage{graphicx}
\usepackage{dcolumn}
\usepackage{bm}
\usepackage{braket}
\usepackage[version=3]{mhchem}
\usepackage{hyperref}
\usepackage{mathrsfs}
\usepackage[version=3]{mhchem}

\newcommand{\equref}[1]{(\ref{#1})}
\newcommand{\figref}[1]{Fig.~\ref{#1}}

\usepackage{color,ulem}

\newcommand{\mrm}{\mathrm}
\newcommand{\mbf}{\mathbf}
\newcommand{\mcl}{\mathcal}

\newcommand{\lt}{\left}
\newcommand{\rt}{\right}

\newcommand{\br}[1]{\left( #1 \right)}

\newcommand{\BR}[1]{\left[ #1 \right]}
\newcommand{\ev}[1]{\left\langle #1 \right\rangle}
\newcommand{\av}[1]{\left| #1 \right|}

\newcommand{\tbra}[1]{\widetilde{\bra{#1}}}
\newcommand{\tket}[1]{\widetilde{\ket{#1}}}

\newcommand{\htab}{\hspace{2em}}
\newcommand{\uar}{\uparrow}
\newcommand{\dar}{\downarrow}

\newcommand{\limto}[1]{\underset{#1}{\longrightarrow}}

\newcommand{\im}{\imath}
\newcommand{\ex}{e}
\newcommand{\alp}{\alpha}

\newcommand{\eps}{\epsilon}

\newcommand{\sig}{\sigma}

\newcommand{\Del}{\Delta}

\newcommand{\veps}{\varepsilon}

\newcommand{\ham}{\mcl{H}}

\newcommand{\pjc}{\mcl{P}}

\newcommand{\pri}{\prime}
\newcommand{\dpri}{{\prime\prime}}

\newcommand{\reff}{\mrm{eff}}
\newcommand{\rtot}{\mrm{tot}}

\newcommand{\rspn}{\mrm{spn}}
\newcommand{\rstr}{\mrm{str}}

\newcommand{\vS}{\bm{S}}
\newcommand{\tS}{\widetilde{S}}
\newcommand{\vtS}{\widetilde{\bm{S}}}
\newcommand{\vchi}{\bm{\chi}}
\newcommand{\tM}{\widetilde{M}}
\newcommand{\thz}{\widetilde{h}_z}

\begin{document}

\title{Quasi-fractionalization of spin in a cluster-based Haldane state\\ supported by chirality of a triangular spin tube}

\author{Takanori Sugimoto}%
\email{sugimoto.takanori@rs.tus.ac.jp}
\affiliation{Department of Applied Physics, Tokyo University of Science, Katsushika, Tokyo 125-8585, Japan}
\author{Takami Tohyama}
\affiliation{Department of Applied Physics, Tokyo University of Science, Katsushika, Tokyo 125-8585, Japan}

\date{\today}

\begin{abstract}
Fractionalization of quantum degrees of freedom holds the key to finding new phenomena in physics, e.g., the quark model in hadron physics, the spin-charge separation in strongly-correlated electron systems, and the fractional quantum Hall effect.
A typical example of the fractionalization in quantum spin systems is the spin-$1$ Haldane state, whose intriguing characteristics are well described by fractionalized $S=1/2$ virtual spins in a bilinear-biquadratic spin-$1$ chain, the so-called Affleck--Kennedy--Lieb--Tasaki model, delivering two individual spin-$1/2$ degrees of freedom as edge states.
Here we theoretically propose an exotic extension of the Haldane state and chirality in a triangular spin tube, inducing a {\it quasi}-fractionalization of spin-$1/2$ degree of freedom, i.e., a {\it quarter} spin.
Existence of the edge state is confirmed both analytically and numerically, combining a low-energy perturbation theory and variational matrix-product state method. 
Our study can not only propose a new quantum spin property but pave a way to novel quantum states of matter.
\end{abstract}

\pacs{Valid PACS appear here}
\maketitle

Historically, developments of physics have often been supported by discoveries of new fractionalization mechanism and derivative hidden degrees of freedom.
Elementary particles in the standard model are the fruits of fractionalization, and moreover, there still remain some fractionalized particles, e.g., a magnetic monopole~\cite{Dirac1931,Dirac1948} and an axion~\cite{Wilczek1987,Bertone2018}, in the dark.
The concept of fractionalization also plays a key role in condensed-matter physics, leading to discoveries of the spin-charge separation in strongly-correlated electron systems~\cite{Lee2006}, the fractional quantum Hall effect~\cite{Stormer1999,Murthy2003}, and the Majorana fermion in topological superconductors~\cite{Qi2011}.
Furthermore, as quantum spin counterparts, intensive studies have been performed on the Haldane state in a spin-$1$ chain~\cite{Haldane1983,Haldane1983,Haldane2017} and the Kitaev model with anisotropic spin interactions~\cite{Kitaev2006} so far.
Particularly, the Haldane state has been used to demonstrate exotic phenomena, symmetry-protected topological phase~\cite{Pollmann2010,Chen2011,Pollmann2012,Chen2012} and holographic quantum computing~\cite{Gross2007,Brennen2008,Miyake2010,Bartlett2010,Else2012}, thanks to its simplicity of both analytical and numerical calculations.
 
\begin{figure}[htpb]
  \centering
  \includegraphics[width=0.85\linewidth]{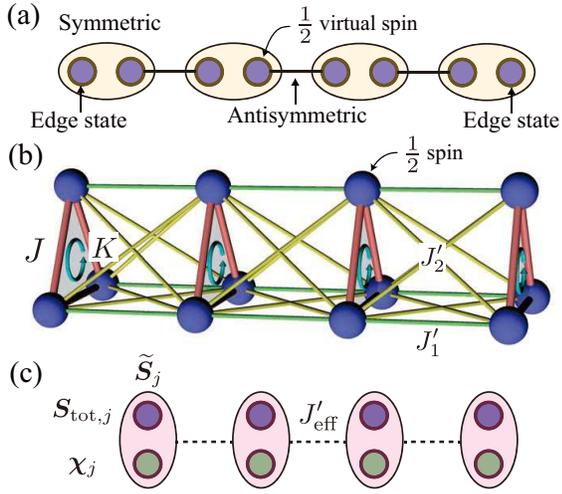}
  \caption{(a) Schematic ground state of the AKLT model. An $S=1$ spin is decomposed into two $S=1/2$ virtual spins (two blue balls in an orange oval). The orange oval (black solid line) represents symmetric (antisymmetric) configuration. As edge states, there are spin-$1/2$ degrees of freedom disconnected from solid bond with open boundary condition. (b) An $S=1/2$ TST. Blue balls denote $S=1/2$ spins. Red, green, and yellow bonds represent $J$, $J_1^\pri$, and $J_2^\pri$ interactions in \equref{ham2} and \equref{ham3}, where the $K$ term is shown as a cyan circular arrow. (c) Low-energy effective model of \equref{ham1}, corresponding to \equref{eham} with $J_1^\pri=J_2^\pri$. Blue (green) balls represent spin-$1/2$ degrees of freedom corresponding to $\vS_{\rtot,j}$ ($\vchi_j$). Pink oval represents a spin-$1$ degree of freedom ($\vtS_j$) given by symmetrization of $\vS_{\rtot,j}$ and $\vchi_j$. Dashed line denotes the effective antiferromagnetic exchange interaction $J_\reff^\pri=J_1^\pri$.}
  \label{fig1}
\end{figure}

Main characteristics in the Haldane state, which is the ground state in an $S=1$ antiferromagnetic spin chain, are clearly explained in the bilinear-biquadratic spin-$1$ chain, the so-called Affleck--Kennedy--Lieb--Tasaki (AKLT) model~\cite{Affleck1988}.
In the AKLT model, an $S=1$ spin is decomposed into two $S=1/2$ virtual spins at each site, and neighboring spins connected with inter-site bonds are antisymmetrized reflecting the antiferromagnetism [see \figref{fig1}(a)].
To restore the $S=1$ spin, corresponding two spins at each site are symmetrized with a projection operator into an even parity space.
Note that in this state, there is a non-trivial topological order defined by a long-ranged string correlation~\cite{DenNijs1989,Tasaki1991,Kennedy1992}.
This procedure can be extended into more than two spins at each site, resulting in the cluster-based Haldane state (CBHS) in a spin cluster chain (SCC)~\cite{Masuda2006,Fujihala2018,Sugimoto2020}.
Meanwhile, with the open boundary condition, it is well known that the original Haldane state exhibits two edge states corresponding to spin-$1/2$ degrees of freedom [see \figref{fig1}(a)].
Similarly, in the CBHS, there should be the edge states, while the edge site contains many $S=1/2$ spins.
Hence, a naive question arises; if we obtain a way to detect only one of the spins at the edges, how does the spin degree of freedom behave in this state, delivering a new type of fractionalization?

To answer this question, in this letter, we consider a CBHS appearing in a triangular spin tube (TST) coupled with chirality degree of freedom [see \figref{fig1}(b)].
In this model, the chirality plays the role of an $S=1/2$ pseudo spin in the low-energy states~\cite{Kawano1997,Luscher2004}. 
If we add a Heisenberg-type coupling between the real and pseudo spins, it can split the four-fold ground-state degeneracy into a triplet and a singlet in a triangle cluster.
The triplet state is regarded as an $S=1$ spin as a result of hybridization of the real and pseudo spins.
Thus, introducing antiferromagnetic interactions between neighboring triangles which are much smaller than the intra-cluster interactions induces a CBHS.
Since the chirality is not directly coupled with the magnetic field, only the real-spin component at edge site can be controlled with applying the magnetic field.

As a concrete description, we propose the model Hamiltonian of $S=1/2$ TST as follows,
\begin{equation}
\ham_0 = \sum_{j=1}^L \ham_{J}^{(j)} + \sum_{j=1}^L \ham_{K}^{(j)} + \sum_{j=1}^{L-1} \ham_{J^\pri}^{(j)} \label{ham1}
\end{equation}
with
\begin{align}
&\ham_{J}^{(j)} = J\sum_{i< i^\pri}\vS_{i,j}\cdot\vS_{i^\pri,j}, \ \ham_{K}^{(j)} = - K\vS_{\rtot,j}\cdot \vchi_j, \label{ham2}\\
&\ham_{J^\pri}^{(j)} = J_1^\pri \sum_{i} \vS_{i,j}\cdot\vS_{i,j+1} + J_2^\pri \sum_{i \neq i^\pri} \vS_{i,j}\cdot\vS_{i^\pri,j+1}, \label{ham3}
\end{align}
where $L$ is the number of clusters.
The local Hamiltonians, $\ham_{J}^{(j)}$, $\ham_{K}^{(j)}$, and $\ham_{J^\pri}^{(j)}$, represent intra-cluster spin interactions of the $j$-th cluster, a spin-chirality interaction of the $j$-th cluster, and inter-cluster spin interactions between the $j$-th and $(j+1)$-th clusters, respectively.
The $S=1/2$ local spin (the total spin) operator in a cluster is denoted by $\vS_{i,j}$ ($\vS_{\rtot,j} = \sum_i \vS_{i,j}$), where $i=1,2,3$ ($j=1,2,\cdots,L$) denotes the site (cluster) index. 
The $S=1/2$ pseudo spin operator corresponding to the scalar chirality is represented by $\vchi_j$ as explained below.
In this model, we consider antiferromagnetic Heisenberg interactions $J>0$ and $J_k^\pri>0$ ($k=1,2$), and a ferromagnetic Heisenberg-type interaction between $\vS_{\rtot,j}$ and $\vchi_j$ ($K>0$).
As the SCC condition, we assume that the inter-cluster interactions $J_k^\pri$ ($k=1,2$) are much smaller than the intra-cluster interactions $J$ and $K$.
It is worthfully noted that intensive studies on TSTs have been performed both theoretically~\cite{Kawano1997,Luscher2004,Okunishi2005,Fouet2006,Sato2007,Nishimoto2008,Sakai2008,Charrier2010,Okamoto2011,Okunishi2012,Zhao2012,Yonaga2015,Alecio2016} and experimentally~\cite{Schnack2004,Manaka2009,Ivanov2010,Manaka2011,Manaka2019,Hagihala2019}, while to the best of our knowledge, the spin-chirality interaction has not been considered so far.
Realizability of this interaction is discussed below.

\begin{figure}[htpb]
  \centering
  \includegraphics[width=0.85\linewidth]{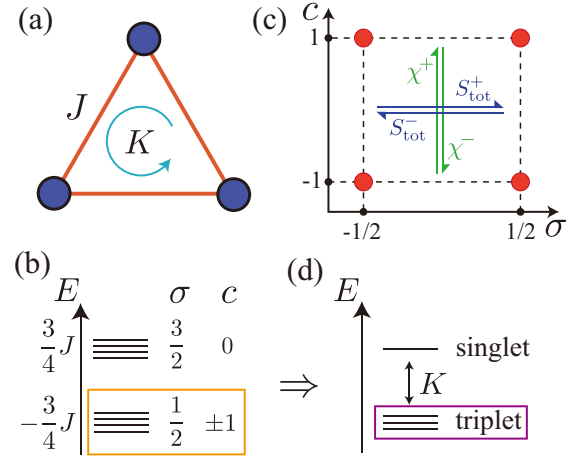}
  \caption{(a) A triangle spin cluster. The $J$ and $K$ terms in \equref{ham2} are denoted by red bond and cyan circular arrow, respectively. (b) Energy spectrum of the triangle cluster with $K=0$, where $\sig$ and $c$ represent the eigenvalues of $S_{\rtot,j}^z$ and $\chi_j$, respectively. There are two quartets. (c) Eigenstates (red balls) of the lower quartet (orangle region) in (b) and ladder operators of $\vS_{\rtot,j}$ and $\vchi_j$. (d) Introducing a finite $K>0$, the lower quartet splits to a singlet and a triplet. The triplet (purple region) is regarded as an $S=1$ pseudo spin.}
  \label{fig2}
\end{figure}

To obtain an effective model based on low-energy perturbation theory, we start with the intra-cluster Hamiltonians \equref{ham2}, shown in \figref{fig2}(a), because of the SCC condition $J_k^\pri \ll J, K$.
Without the $K$ term, the local Hamiltonian corresponds to a uniform antiferromagnetic Heisenberg model on a triangle cluster.
The eigenstates are decomposed into a quartet of $S_{\rtot,j}=3/2$ and degenerate two doublets of $S_{\rtot,j}=1/2$ in \figref{fig2}(b).
All the local eigenstates $\ket{\sig,c}$ are classified by $z$ component of the total spin operator $S_{\rtot,j}^z$ and the scalar chirality operator $\chi_j\equiv (4/\sqrt{3})\vS_{1,j}\cdot\vS_{2,j}\times\vS_{3,j}$, where $\sig$ ($c$) is the eigenvalue of $S_{\rtot,j}^z$ ($\chi_j$), because of the commutation relations, $[\ham_{J}^{(j)},S_{\rtot,j}^z]=[\ham_{J}^{(j)},\chi_j]=[S_{\rtot,j}^z,\chi_j]=0$:
\begin{align}
  &\ket{\pm\frac{3}{2},0}_j = \ket{\pm\pm\pm}_j,\\
  &\ket{\pm\frac{1}{2},0}_j = \frac{\pm 1}{\sqrt{3}}\br{\ket{\pm\pm\mp}_j+\ket{\pm\mp\pm}_j+\ket{\mp\pm\pm}_j}, \\
  &\ket{\pm\frac{1}{2},1}_j = \frac{\pm 1}{\sqrt{3}}\br{\ex^{-\im\phi}\ket{\pm\pm\mp}_j+\ex^{\im\phi}\ket{\pm\mp\pm}_j+\ket{\mp\pm\pm}_j}, \\
  &\ket{\pm\frac{1}{2},-1}_j = \frac{\pm 1}{\sqrt{3}}\br{\ex^{\im\phi}\ket{\pm\pm\mp}_j+\ex^{-\im\phi}\ket{\pm\mp\pm}_j+\ket{\mp\pm\pm}_j},
\end{align}
with $\phi=2\pi/3$ and the imaginary unit $\im=\sqrt{-1}$. The ket states in the right-hand side denote the direct products of one-spin eigenstate, e.g., $\ket{+-+}_j=\ket{\uar}_{1,j}\ket{\dar}_{2,j}\ket{\uar}_{3,j}$. 
Figure 2(c) shows the lower quartet [orange region in \figref{fig2}(b)] in the parameter space of $\sig$ and $c$.
In the same manner as $S=1/2$ spin operators, we can construct a pseudo spin operator $\vchi_j$ with $\chi_j^z\equiv \chi_j/2$ and corresponding ladder operators, $\chi_j^\pm\equiv\sum_{\sig=\pm\frac{1}{2}} \ket{\sig,\pm 1}_j\bra{\sig,\mp 1}_j$.
Hence, the $x$ and $y$ components of the pseudo spin operator are given by $\chi_j^x\equiv(\chi_j^++\chi_j^-)/2$ and $\chi_j^y\equiv(\chi_j^+-\chi_j^-)/(2\im)$, respectively.
Introducing a finite $K>0$ with the pseudo spin operator $\vchi_j$ in $\ham_K^{(j)}$, we can obtain a triplet $\tket{m}_j$ ($m=0,\pm 1$) as local ground states [see \figref{fig2}(d)] given by, 
\begin{align}
  \tket{\pm 1}_j=\ket{\pm\frac{1}{2},\pm 1}_j,\ \tket{0}_j=\frac{1}{\sqrt{2}}\br{\ket{\frac{1}{2},-1}_j+\ket{-\frac{1}{2},1}_j}, 
\end{align}
where $m$ corresponds to the eigenvalue of $\tS_j^z=S_{\rtot,j}^z+\chi_j^z$.
Note that the $K$ term does not affect the upper quartet, because the eigenvalue of the chirality is zero ($c=0$) in the upper quartet.

At low temperatures $T\lesssim K$, only the local triplet states $\tket{m}_j$ are relevant to low-energy physics with small enough inter-cluster interactions.
This corresponds to the $S=1$ pseudo spin, $\vtS_j\equiv\pjc_j(\vS_{\rtot,j}^z+\vchi_j^z)\pjc_j$ with a projection operator $\pjc_j=\sum_m \tket{m}_j\tbra{m}_j$.
Moreover, we can obtain an effective Hamiltonian of the pseudo spin-$1$ operator (see Appendix~A for the derivation) as follows,
\begin{widetext}
\begin{align}
  \ham_{\reff}&=\sum_j \Bigg\{\BR{\br{\frac{5}{3}J_1^\pri-\frac{2}{3}J_2^\pri}\tS_j^z\tS_{j+1}^z+J_1^\pri\br{\tS_j^x\tS_{j+1}^x+\tS_j^y\tS_{j+1}^y}} + \frac{8}{3}(J_1^\pri-J_2^\pri)\br{ \frac{1}{2}\tS_j^z\tS_{j+1}^z+\tS_j^x\tS_{j+1}^x+\tS_j^y\tS_{j+1}^y}^2 \notag\\
 &\hspace{3em} -\frac{2}{3}(J_1^\pri-J_2^\pri)\BR{2-(\tS_j^z)^2}\BR{2-(\tS_{j+1}^z)^2} \Bigg\} +\mrm{const}. \label{eham}
\end{align}
\end{widetext}
In the effective Hamiltonian, there are an anisotropic exchange interaction, an anisotropic biquadratic interaction, and an additional term favoring zero magnetization.
Assuming $J_1^\pri=J_2^\pri$, the effective Hamiltonian is equivalent to a Heisenberg chain of the $S=1$ pseudo spins.
Therefore, the ground state of the SCC Hamiltonian \equref{ham1} is a CBHS consisting of the $S=1$ pseudo spins with $J_1^\pri=J_2^\pri$.


\begin{figure}[htpb]
  \centering
  \includegraphics[width=0.9\linewidth]{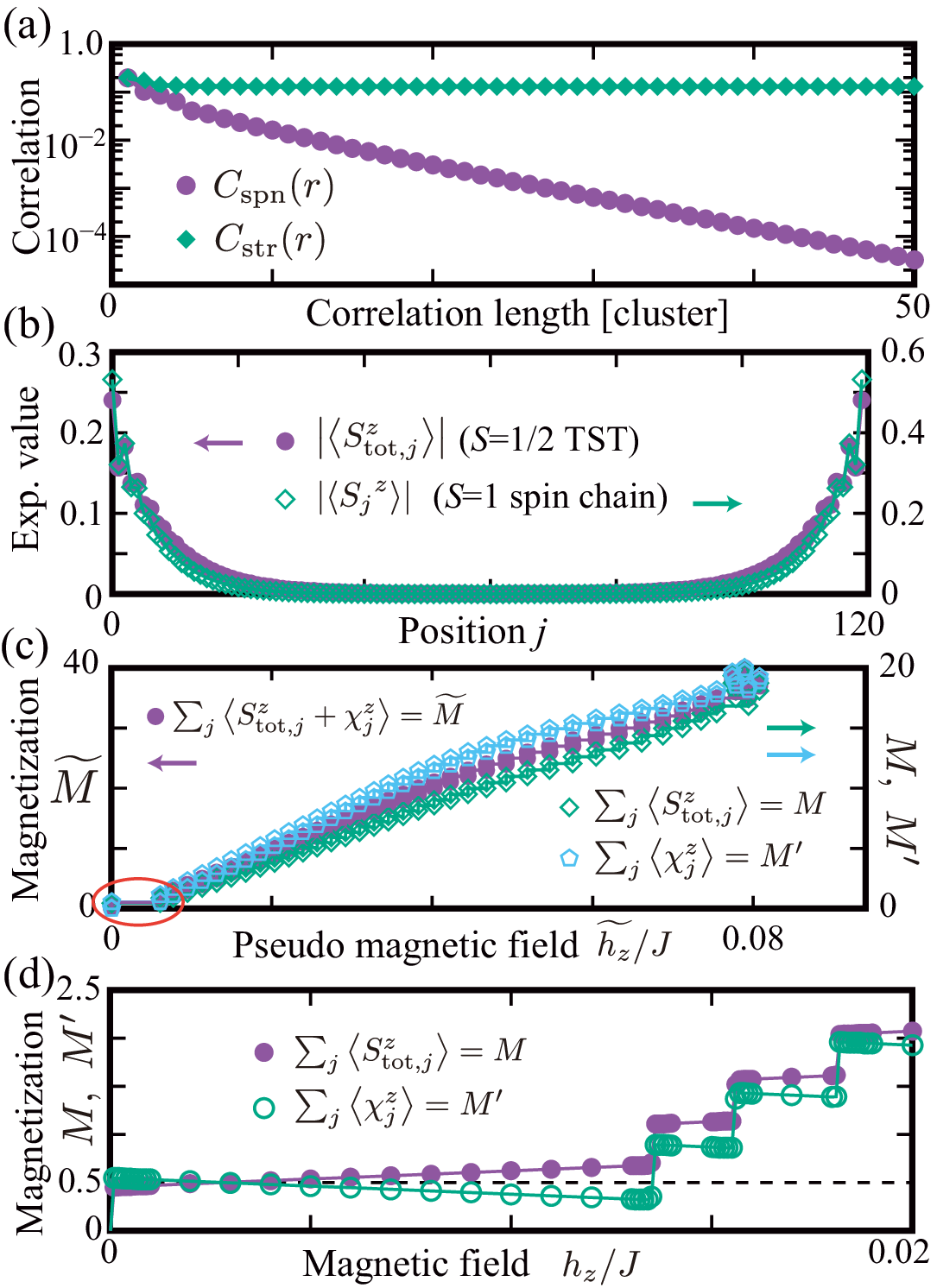}
  \caption{(a) Pseudo spin and string correlation functions $C_\rspn(r)$ and $C_\rstr(r)$ at the $\tM=0$ ground state in an $L=120$ ($D=500$) TST. (b) Local absolute expectation value of the real spin $|\ev{S_{\rtot,j}^z}|$ at the $\tM=1$ eigenstate in an $L=120$ ($D=500$) TST. For comparison, expectation value of spin $|\ev{S_j^z}|$ is shown at $M=1$ in an $N=120$ ($D=300$) spin-$1$ chain. (c) (Pseudo) magnetization $\tM$, $M^\pri$, and $M$ curves with applied pseudo magnetic field $\thz$ in an $L=40$ ($D=300$) TST. Red oval denotes a magnetization plateau at $\tM=1$. (d) Response of expectation values of the real and pseudo spins $\sum_j\ev{S_{\rtot,j}^z}$ and $\sum_j\ev{\chi_j^z}$ to real magnetic field $h_z$ in an $L=60$ ($D=300$) TST. The dashed line denotes the $M$ ($M^\pri$) $=0.5$ level.}
  \label{fig3}
\end{figure}

To confirm this statement, we have performed numerical calculations for TST \equref{ham1} by the variational matrix-product state method~\cite{Schollwock2011} (see Appendix~B).
In the calculations, we have obtained several expectation values and correlation functions in addition to eigenstates and eigen-energies.
We have checked the sufficient convergence (the truncation error $\eps_{\mrm{trunc}}\lesssim 10^{-6}$) for the bond dimension $D\geq 300$.
As the parameters of Hamiltonians, we choose $J_1^\pri=J_2^\pri=J/(10\sqrt{2})$ and $K=J/2$ for the results in \figref{fig3}.

Firstly, we show pseudo spin and string correlation functions~\cite{DenNijs1989,Tasaki1991,Kennedy1992} defined by,
\begin{align}
  C_{\rspn}(r)&=\ev{\tS_j^z\tS_{j+r}^z},\\
  C_{\rstr}(r)&=\ev{\tS_{j}^z\exp\br{\im \pi \sum_{k=j+1}^{j+r-1} \tS_k^z } \tS_{j+r}^z}.
\end{align}
For the numerical calculation, we have chosen $j=L/2-\lt\lfloor r/2\rt\rfloor$, where the floor function $\lfloor x\rfloor$ represents the integer part of $x$. 
Figure 3(a) shows the correlation functions at the ground state in an $L=120$ TST \equref{ham1}.
We can see convergence of the string correlation to a finite constant irrespective to the exponential decay of the spin correlation, indicating the (cluster-based) Haldane state.

Next, we check existence of edge states in $\tM=\sum_j\ev{\tS_j^z}=1$ eigenstate, which is the first excited state with an energy gap converging to zero in the thermodynamical limit $\Del_1\limto{L\to\infty}0$.
In \figref{fig3}(b), two expectation values in absolute value are shown: $\av{\ev{S_{\rtot,j}^z}}$ in an $L=120$ TST and for comparison, $\av{\ev{S_j^z}}$ in an $N=120$ spin-$1$ chain, where $N$ is the number of spins, exhibiting edge states in the Haldane state.
The pseudo spin's edge states in the TST are localized and decoupled at two edges as well as the edge states of the Haldane state.
The real-spin component $\ev{S_{\rtot,j}^z}$ is similarly distributed, while the absolute value is almost the half of the spin $\av{\ev{S_j^z}}$ in the Haldane state.
The edge state in the original Haldane state has the magnitude of $S=1/2$ spin, so that the real-spin component $\ev{S_{\rtot,j}^z}$ in the TST can be regarded as a half of the $S=1/2$ spin, i.e., a {\it quarter} ($S=1/4$) spin.
It is not the case that the real-spin component does neither obey any new algebra nor have any new group of an $S=1/4$ spin, but rather the case that the magnitude is almost equivalent to a half of $S=1/2$ spin.

Moreover, to comfirm stability of the edge states, we examine response to two types of external fields given by,
\begin{align}
  \ham_{Z} &= -h_z\sum_j S_{\rtot,j}^z,\ \widetilde{\ham}_{Z} = -\thz\sum_j (S_{\rtot,j}^z+\chi_j^z).
\end{align}
The former (latter) Halmitonian is the Zeeman term of real (pseudo) spins.
Note that the TST Hamiltonian \equref{ham1} does not commute with the former term, but the latter term, because the Heisenberg-type interaction of the real spin and chirality $\vS_{\rtot,j}\cdot\vchi$ in $\ham_K^{(j)}$ breaks the conservation law of total spin and keeps the sum of total spin and chirality. 
Therefore, instead of the real magnetization $M=\sum_j \ev{S_{\rtot,j}^z}$, pseudo magnetization $\tM=\sum_j \ev{\tS_j^z}$ is a good quantum number.
Figure 3(c) presents the pseudo magnetization curves with applied the pseudo magnetic field $\thz$, whereas the real magnetization and chirality versus the real magnetic field $h_z$ are shown in \figref{fig3}(d).
In the pseudo magnetization [\figref{fig3}(c)], we can see the zero-magnetization plateau with $\tM=1$ edge states corresponding to the Haldane gap between $\tM=1$ and $2$.
Interestingly, with applying a small real magnetic field in \figref{fig3}(d), a quasi-plateau with a small tilt appears in both the real magnetization $M$ and chirality $M^\pri$.
The expectation value of real magnetization $M\cong 1/2$ is composed of two decoupled edge states like \figref{fig3}(b).
Thus, the quarter-spin magnetization can be observed at the edges even with a real magnetic field.

Lastly, we comment on a possibility of experimental setup of the TST model.
In our model, the Heisenberg terms in a TST, i.e., $\ham_J^{(j)}$ and $\ham_{J^\pri}^{(j)}$, are familiar in many quantum spin materials.
Hence, the coupling term of the real spin and the chirality  $\ham_K^{(j)}$ is only distinct in light of real compounds.
To understand this term in detail, we rewrite it in original spin operators, leading to
\begin{align}
  &S_{\rtot,j}^z \chi_j^z = \frac{1}{4\sqrt{3}}\sum_{i,i^\pri,i^\dpri} \eps_{ii^\pri i^\dpri} (\vS_{i^\pri,j} \times \vS_{i^\dpri,j})^z, \label{Kzz} \\
  &S_{\rtot,j}^+ \chi_j^-+\mrm{H.c.} = -\pjc_j^d(S_{1,j}^x+S_{2,j}^{-\phi}+S_{3,j}^\phi)\pjc_j^d. \label{Kxy}
\end{align}
Here, we use the projection operator into the cluster's doublets $\pjc_{j}^d=(15/4-\vS_{\rtot,j}^2)/3$ and $\phi$ component of spin $S_{i,j}^{\pm \phi}=(\ex^{\mp \im \phi}S_{i,j}^++\ex^{\pm \im \phi}S_{i,j}^-)/2$ with $\phi=2\pi/3$. 
Apparently, the Ising part of this interaction \equref{Kzz} corresponds to $z$ component of the vector chirality.
Meanwhile, the XY part \equref{Kxy} is regarded as a transverse herical magnetic field with the projection into doublets in a cluster.
The projection can be effectively introduced at low temperatures if we consider the case $J\gg K$, although the triplet is slightly broken with a small energy gap $\veps\sim K^2/J$ due to the hybridization effects of the quartet and doublets in a cluster.
Note that in this case, the intra-cluster interactions need to be approximately greater than the energy gap $J_k^\pri \gtrsim K^2/J$, to exhibit the CBHS.
Therefore, this term may be found in a chiral magnet holding a vector chirality, with applied the transverse herical magnetic field at low temperatures.

In conclusion, we have proposed an exotic extension of the Haldane state to show a novel mechanism of fractionalization of edge states.
In this model, the chirality degree of freedom is regarded as an $S=1/2$ pseudo spin.
The real and pseudo spins are symmetrized in the Haldane state of our model, so that an $S=1/2$ spin degree of freedom appearing as the edge states consists of the real and pseudo spin components.
Since the magnetic field is directly coupled with only the real spin, we can observe a half of the magnitude of edge states, corresponding to a quarter spin.
Our concept not only gives a new quantum spin feature, but also implies various possibilities of quantum fractionalization.

\begin{acknowledgments}
We would like to thank M. Fujihala, S. Mitsuda, and K. Morita for giving us a motivation of this study through the preceding studies of Fedotovite and a Kagome strip. 
This work was partly supported by a Grant-in-Aid for Young Scientists (B) (Grant No.~16K17753), Grant-in-Aid for Scientific Research (C) (Grant No.~20K03840).
Numerical computation in this work was carried out on the supercomputers at JAEA and the Supercomputer Center at the Institute for Solid State Physics, University of Tokyo.
\end{acknowledgments}

\appendix
\section{Derivation the effective Hamiltonian}
In this section, we show an explicit derivation of the effective Hamiltonian from the original Hamiltonian of a triangular spin tube:
\begin{equation}
\ham_0 = \sum_{j=1}^L \ham_{J}^{(j)} + \sum_{j=1}^{L-1} \ham_{J^\pri}^{(j)} + \sum_{j=1}^L \ham_{K}^{(j)} \label{ham1}
\end{equation}
with
\begin{align}
&\ham_{J}^{(j)} = J\sum_{i< i^\pri}\vS_{i,j}\cdot\vS_{i^\pri,j}, \label{ham2} \\ 
&\ham_{J^\pri}^{(j)} = J_1^\pri \sum_{i} \vS_{i,j}\cdot\vS_{i,j+1} + J_2^\pri \sum_{i \neq i^\pri} \vS_{i,j}\cdot\vS_{i^\pri,j+1}, \label{ham3} \\
&\ham_{K}^{(j)} = - K\vS_{\rtot,j}\cdot \vchi_j, \label{ham4}
\end{align}
where $L$ is the number of clusters.
The local Hamiltonians, $\ham_{J}^{(j)}$, $\ham_{J^\pri}^{(j)}$, and $\ham_{K}^{(j)}$, represent intra-cluster spin interactions of the $j$-th cluster, inter-cluster spin interactions between the $j$-th and $(j+1)$-th clusters, and a spin-chirality interaction of the $j$-th cluster, respectively.
The $S=1/2$ local spin (the total spin) operator in a cluster is denoted by $\vS_{i,j}$ ($\vS_{\rtot,j} = \sum_i \vS_{i,j}$), where $i=1,2,3$ ($j=1,2,\cdots,L$) denotes the site (cluster) index. 
The $S=1/2$ pseudo spin operator $\vchi_j=(\chi_j^x,\chi_j^y,\chi_j^z)$ based on the scalar chirality $\chi_j\equiv (4/\sqrt{3})\vS_{1,j}\cdot\vS_{2,j}\times\vS_{3,j}$ is defined by,
\begin{align}
  \chi_j^x=(\chi_j^++\chi_j^-)/2,\ \chi_j^y=(\chi_j^+-\chi_j^-)/(2\im),\ \chi_j^z= \chi_j/2
\end{align}
with the chirality ladder operators $\chi_j^\pm\equiv\sum_{\sig=\pm\frac{1}{2}} \ket{\sig,\pm 1}_j\bra{\sig,\mp 1}_j$ and the imaginary unit $\im=\sqrt{-1}$. Here, $\ket{\sig,c}$ ($\sig=\pm 1/2$ and $c=\pm 1$) represents the simultaneous eigenstates of the total spin $S_{\rtot,j}^z$ and the scalar chirality.

The triplet ground states with $J>0$ and $K>0$ in the cluster Hamiltonians \equref{ham2} and \equref{ham4} are given by,
\begin{align}
  &\tket{1}_j=\frac{1}{\sqrt{3}}\br{\ex^{-\im\phi}\ket{\uar\uar\dar}_j+\ex^{\im\phi}\ket{\uar\dar\uar}_j+\ket{\dar\uar\uar}_j},\\
  &\tket{0}_j=\frac{1}{\sqrt{6}}\big( \ex^{\im\phi}\ket{\uar\uar\dar}_j+\ex^{-\im\phi}\ket{\uar\dar\uar}_j+\ket{\dar\uar\uar}_j \notag\\
    & \htab\htab - \ex^{-\im\phi}\ket{\dar\dar\uar}_j-\ex^{\im\phi}\ket{\dar\uar\dar}_j-\ket{\uar\dar\dar}_j\big), \\
  &\tket{-1}_j=-\frac{1}{\sqrt{3}}\br{\ex^{\im\phi}\ket{\dar\dar\uar}_j+\ex^{-\im\phi}\ket{\dar\uar\dar}_j+\ket{\uar\dar\dar}_j},
\end{align}
with $\phi=2\pi/3$. The ket states in the right-hand side denote the direct products of one-spin eigenstates, e.g., $\ket{\uar\dar\uar}_j=\ket{\uar}_{1,j}\ket{\dar}_{2,j}\ket{\uar}_{3,j}$. 
By using the triplet states, the projection operator is defined by $\pjc_j=\sum_{m=0,\pm 1} \tket{m}_j\tbra{m}_j$.

At low temperatures $T\ll J, K$, the low-energy physics is well described by the triplet states, neglecting the other states of cluster. 
Since the triplet states correspond to the eigenstates of an $S=1$ spin, we can write projected operators of the local spins with the $S=1$ pseudo spin operator $\vtS_j$,
\begin{widetext}
\begin{align}
  &\pjc_j\vS_{1,j}\pjc_j = \pjc_j \begin{pmatrix} S_{1,j}^x \\ S_{1,j}^y \\ S_{1,j}^z \end{pmatrix} \pjc_j = \frac{1}{3}
  \begin{pmatrix}
    \tS_j^x - (\tS_j^+)^2-(\tS_j^-)^2 + 2(\tS_j^z)^2 -2 \\
    \tS_j^y + \im (\tS_j^+)^2 - \im(\tS_j^-)^2 \\
    \tS_j^z - \tS_j^z\tS^+ - \tS_j^z\tS^- - \tS_j^+\tS_j^z -\tS_j^-\tS_j^z
  \end{pmatrix},\label{eS1} \\
  &\pjc_j\vS_{2,j}\pjc_j = \pjc_j \begin{pmatrix} S_{2,j}^x \\ S_{2,j}^y \\ S_{2,j}^z \end{pmatrix} \pjc_j = \frac{1}{3}
  \begin{pmatrix}
    \tS_j^x - \ex^{\im\phi} (\tS_j^+)^2 - \ex^{-\im\phi} (\tS_j^-)^2 - (\tS_j^z)^2 +1 \\
    \tS_j^y - \ex^{\im\phi/4} (\tS_j^+)^2 - \ex^{-\im\phi/4} (\tS_j^-)^2 -\sqrt{3}(\tS_j^z)^2 +\sqrt{3} \\
    \tS_j^z - \ex^{\im\phi}\tS_j^z\tS^+ - \ex^{-\im\phi} \tS_j^z\tS^- - \ex^{\im\phi} \tS_j^+\tS_j^z - \ex^{-\im\phi} \tS_j^-\tS_j^z
  \end{pmatrix},\label{eS2}\\
  &\pjc_j\vS_{3,j}\pjc_j = \pjc_j \begin{pmatrix} S_{3,j}^x \\ S_{3,j}^y \\ S_{3,j}^z \end{pmatrix} \pjc_j = \frac{1}{3}
  \begin{pmatrix}
    \tS_j^x - \ex^{-\im\phi} (\tS_j^+)^2 - \ex^{\im\phi} (\tS_j^-)^2 - (\tS_j^z)^2 +1 \\
    \tS_j^y + \ex^{-\im\phi/4} (\tS_j^+)^2 + \ex^{\im\phi/4} (\tS_j^-)^2 +\sqrt{3}(\tS_j^z)^2 -\sqrt{3} \\
    \tS_j^z - \ex^{-\im\phi}\tS_j^z\tS^+ - \ex^{\im\phi} \tS_j^z\tS^- - \ex^{-\im\phi} \tS_j^+\tS_j^z - \ex^{\im\phi} \tS_j^-\tS_j^z
  \end{pmatrix} \label{eS3}.
\end{align}
\end{widetext}
Although it is not so easy to calculate the inter-cluster interactions step by step with these operators, the total spin has a simple form $\pjc_j\vS_{\rtot,j}\pjc_j = \vtS_j$.
Therefore, we can easily obtain the effective Hamiltonian of an equivalent case of the inter-cluster interactions $J_1^\pri=J_2^\pri(\equiv J^\pri)$,
\begin{align}
  &\ham_{\reff}^{(j)}|_{J_1^\pri=J_2^\pri=J^\pri}= (\pjc_j\pjc_{j+1})\ham_{J^\pri}^{(j)}|_{J_1^\pri=J_2^\pri}(\pjc_j\pjc_{j+1})\notag\\
&\htab = J^\pri (\pjc_j\pjc_{j+1})(\vS_{\rtot,j}\cdot\vS_{\rtot,j+1})(\pjc_j\pjc_{j+1})=J^\pri\vtS_j\cdot\vtS_{j+1}. \label{hami}
\end{align}
On the other hand, though a discord case $J_1^\pri\neq J_2^\pri$ is not so easy, the case of $J_2^\pri=0$ is relatively easy to obtain.
Moreover, if the effective Hamiltonian of (i) $J_1^\pri\neq 0$ and $J_2^\pri=0$ is obtained, we can also obtain the effective Hamiltonian for (ii) $J_1^\pri= 0$ and $J_2^\pri\neq 0$, because the equivalent case of effective Hamiltonian \equref{hami} is the sum of both the cases,
\begin{equation}
  \ham_{\reff}^{(j)}|_{J_1^\pri=J_2^\pri=J^\pri} = \ham_{\reff}^{(j)}|_{J_1^\pri= J^\pri\neq 0, J_2^\pri=0} + \ham_{\reff}^{(j)}|_{J_1^\pri= 0, J_2^\pri=J^\pri \neq 0}.
\end{equation}
Then, to obtain the effective Hamiltonian in general case, it is sufficient to show the case of (i) $J_1^\pri\neq 0$ and $J_2^\pri=0$.
With the relation $\sum_i\br{\pjc_j\vS_{i,j}\pjc_j - \frac{1}{3}\vtS_j}=0$, the effective Hamiltonian of the $J_1^\pri$ term is rewritten by,
\begin{widetext}
\begin{align}
  \ham_{\reff}^{(j)}|_{J_1^\pri\neq 0, J_2^\pri=0}&= J_1^\pri \sum_{\alp,i}(\pjc_j\pjc_{j+1})(S_{i,j}^\alp S_{i,j+1}^\alp)(\pjc_j\pjc_{j+1})\notag\\
 &= J_1^\pri \sum_{\alp,i}\BR{\br{\pjc_j S_{i,j}^\alp \pjc_{j}-\frac{1}{3}\tS_j^\alp}+\frac{1}{3}\tS_j^\alp}\BR{\br{\pjc_{j+1}S_{i,j+1}^\alp\pjc_{j+1}-\frac{1}{3}\tS_{j+1}^\alp}+\frac{1}{3}\tS_{j+1}^\alp}\notag\\
&= \frac{J_1^\pri}{3} \vtS_j\cdot\vtS_{j+1} + J_1^\pri\sum_{\alp,i}\br{\pjc_j S_{i,j}^\alp \pjc_{j}-\frac{1}{3}\tS_j^\alp}\br{\pjc_{j+1}S_{i,j+1}^\alp\pjc_{j+1}-\frac{1}{3}\tS_{j+1}^\alp}.
\end{align}
\end{widetext}
According to \equref{eS1}--\equref{eS3}, we can rewrite the second term into
\begin{align}
 &\sum_{\alp,i}\br{\pjc_j S_{i,j}^\alp \pjc_{j}-\frac{1}{3}\tS_j^\alp}\br{\pjc_{j+1}S_{i,j+1}^\alp\pjc_{j+1}-\frac{1}{3}\tS_{j+1}^\alp} \notag\\
&\htab =\frac{1}{9}\sum_{\alp,i} \sum_{m,n}(a_{i,m}^\alp O_{j,m}^\alp)(a_{i,n}^\alp O_{j+1,n}^\alp) \notag\\
&\htab =\frac{1}{9}\sum_{\alp} \sum_{m,n} O_{j,m}^\alp\br{\sum_i a_{i,m}^\alp a_{i,n}^\alp} O_{j+1,n}^\alp \label{covec}, 
\end{align}
with coefficient and operator vectors, $\bm{a}_i^\alp=\{a_{i,m}^\alp\}$ and $\bm{O}_j^\alp=\{O_{j,m}^\alp\}$ ($m=1,2,3,4$) given by,
\begin{align}
  &\bm{a}_1^x=-\begin{pmatrix} 1, & 1, & -2, & 2\end{pmatrix},\\
  &\bm{a}_2^x=-\begin{pmatrix} \ex^{\im\phi}, & \ex^{-\im\phi}, & 1, & -1\end{pmatrix},\\
  &\bm{a}_3^x=-\begin{pmatrix} \ex^{-\im\phi}, & \ex^{\im\phi}, & 1, & -1\end{pmatrix},\\
  &\bm{a}_1^y=\begin{pmatrix} \im, & -\im, & 0, & 0\end{pmatrix},\\
  &\bm{a}_2^y=-\begin{pmatrix} \ex^{\im\phi/4}, & \ex^{-\im\phi/4}, & \sqrt{3}, & -\sqrt{3}\end{pmatrix},\\
  &\bm{a}_3^y=\begin{pmatrix} \ex^{-\im\phi/4}, & \ex^{\im\phi/4}, & \sqrt{3}, & -\sqrt{3}\end{pmatrix},\\
  &\bm{a}_1^z=-\begin{pmatrix} 1, & 1, & 1, & 1\end{pmatrix},\\
  &\bm{a}_2^z=-\begin{pmatrix} \ex^{\im\phi}, & \ex^{-\im\phi}, & \ex^{\im\phi}, & \ex^{-\im\phi}\end{pmatrix},\\
  &\bm{a}_3^z=-\begin{pmatrix} \ex^{-\im\phi}, & \ex^{\im\phi}, & \ex^{-\im\phi}, & \ex^{\im\phi}\end{pmatrix},
\end{align}
and
\begin{align}
  &\bm{O}_j^x=\bm{O}_j^y=\begin{pmatrix} (\tS_j^+)^2, & (\tS_j^-)^2, & (\tS_j^z)^2, & 1\end{pmatrix},\\
  &\bm{O}_j^z=\begin{pmatrix} \tS_j^z\tS_j^+, & \tS_j^z\tS_j^-, & \tS_j^+\tS_j^z, & \tS_j^-\tS_j^z\end{pmatrix}.
\end{align}
The coefficient matrices in \equref{covec} are obtained as the direct product of the coefficient vectors, 
\begin{equation}
 \mbf{A}_i^\alp=\{(A_i^\alp)_{m,n} \}= \{a_{i,m}^\alp a_{i,n}^\alp\} = \bm{a}_i^x \otimes \bm{a}_i^x.
\end{equation}
The sum of the coefficient matrices $\mbf{A}_{\rtot}^\alp=\sum_i\mbf{A}_i^\alp$ is easily calculated, leading to
\begin{align}
  &\mbf{A}_{\rtot}^x =
  \begin{pmatrix}
    0 & 3 & -3 & 3\\
    3 & 0 & -3 & 3\\
    -3 & -3 & 6 & -6\\
    3 & 3 & -6 & 6
  \end{pmatrix},\\
  &\mbf{A}_{\rtot}^y =
  \begin{pmatrix}
    0 & 3 & 3 & -3\\
    3 & 0 & 3 & -3\\
    3 & 3 & 6 & -6\\
    -3 & -3 & -6 & 6
  \end{pmatrix},\\
  &\mbf{A}_{\rtot}^z =
  \begin{pmatrix}
    0 & 3 & 0 & 3\\
    3 & 0 & 3 & 0\\
    0 & 3 & 0 & 3\\
    3 & 0 & 3 & 0
  \end{pmatrix}.
\end{align}
With these matrices, we can rewrite \equref{covec} into
\begin{align}
  &\sum_{\alp,i}\br{\pjc_j S_{i,j}^\alp \pjc_{j}-\frac{1}{3}\tS_j^\alp}\br{\pjc_{j+1}S_{i,j+1}^\alp\pjc_{j+1}-\frac{1}{3}\tS_{j+1}^\alp} \notag\\
  &\htab = \frac{J_1^\pri}{9}\BR{\bm{O}_{j}^x (\mbf{A}_{\rtot}^x+\mbf{A}_{\rtot}^y) (\bm{O}_{j}^x)^T +\bm{O}_{j}^z \mbf{A}_{\rtot}^z (\bm{O}_{j}^z)^T}.
\end{align}
Calculating the vector-matrix-vector products in the right-hand side, we finally obtain the effective Hamiltonian of (i) $J_1^\pri\neq 0$ and $J_2^\pri=0$ as follows,
\begin{widetext}
\begin{align}
  \ham_{\reff}^{(j)}|_{J_1^\pri\neq 0, J_2^\pri=0}&=J_1^\pri \Bigg\{\br{\frac{5}{3} \tS_j^z\tS_{j+1}^z + \tS_j^x\tS_{j+1}^x + \tS_j^y\tS_{j+1}^y }+\frac{8}{3}\br{\frac{1}{2}\tS_j^z\tS_{j+1}^z + \tS_j^x\tS_{j+1}^x + \tS_j^y\tS_{j+1}^y }^2\notag\\
 &\hspace{4em} -\frac{2}{3} \BR{2-(\tS_j^z)^s}\BR{2-(\tS_{j+1}^z)^s} -\frac{4}{3}\Bigg\}.
\end{align}
Thus, the effective Hamiltonian of (ii) $J_1^\pri= 0$ and $J_2^\pri\neq 0$ is obtained by,
\begin{align}
 &\ham_{\reff}^{(j)}|_{J_1^\pri= 0, J_2^\pri\neq 0} = \frac{J_2^\pri}{J^\pri} \br{\ham_{\reff}^{(j)}|_{J_1^\pri\to J^\pri, J_2^\pri\to J^\pri} - \ham_{\reff}^{(j)}|_{J_1^\pri\to J^\pri, J_2^\pri=0} } \notag\\
& \hspace{3em}=-J_2^\pri \Bigg\{\frac{2}{3} \tS_j^z\tS_{j+1}^z +\frac{8}{3}\br{\frac{1}{2}\tS_j^z\tS_{j+1}^z - \tS_j^x\tS_{j+1}^x + \tS_j^y\tS_{j+1}^y }^2+\frac{2}{3} \BR{2-(\tS_j^z)^s}\BR{2-(\tS_{j+1}^z)^s} -\frac{4}{3}\Bigg\}.
\end{align}
\end{widetext}
We have also confirmed the derivation of effective Hamiltonians from the original Hamiltonians with the matrix form in the two-cluster Hilbert spaces, corresponding to the projection of $(2^3)^2\times (2^3)^2$ matrices to $3^2 \times 3^2$ matrices.

\section{Matrix-product operator representations of the Hamiltonian}
In this section, we show the matrix-product operator (MPO) representation of the Hamiltonian, which is used in the variational matrix-product state (VMPS) method.
For simplicity, we firstly devide the Hamiltonian \equref{ham1} into two-body terms preserving the magnetization and three-body terms (including one-body terms) breaking the magnetization,
\begin{equation}
  \ham_0=\ham_{2b} + \ham_{3b}
\end{equation}
with
\begin{widetext}
\begin{align}
  \ham_{2b}&=\sum_{j=1}^L \ham_{J}^{(j)} + \sum_{j=1}^{L-1} \ham_{J^\pri}^{(j)} -K \sum_{j=1}^L S_{\rtot,j}^z\chi_j^z \notag\\
  &=J\sum_{j=1}^L \sum_{i< i^\pri}\vS_{i,j}\cdot\vS_{i^\pri,j} + J_1^\pri\sum_{j=1}^{L-1} \sum_{i} \vS_{i,j}\cdot\vS_{i,j+1} + J_2^\pri \sum_{j=1}^{L-1} \sum_{i \neq i^\pri} \vS_{i,j}\cdot\vS_{i^\pri,j+1} -\frac{K}{4\sqrt{3}} \sum_{j=1}^L \sum_{i,i^\pri,i^\dpri} \eps_{i i^\pri i^\dpri}(\vS_{i^\pri,j}\times\vS_{i^\dpri,j})^z\\
  \ham_{3b}&= -\frac{K}{2} \sum_{j=1}^L (S_{\rtot,j}^+\chi_j^- +S_{\rtot,j}^-\chi_j^+) =  -\frac{2K}{3} \sum_{j=1}^L\BR{S_1^x\, \mrm{Hs}_j(2,3)+S_2^\phi\, \mrm{Hs}_j(3,1)+S_3^{-\phi}\, \mrm{Hs}_j(1,2)} \label{ham3b}.
\end{align}
\end{widetext}
Here, we use the energy-shifted Heisenberg interaction in the $j$-th cluster $\mrm{Hs}_j(i,i^\pri)=\vS_{i,j}\cdot\vS_{i^\pri,j}-1/4$, and $\phi$ component of spin $S_{i,j}^{\pm \phi}=(\ex^{\mp \im \phi}S_{i,j}^++\ex^{\pm \im \phi}S_{i,j}^-)/2$ with $\phi=2\pi/3$.

According to a review of the VMPS method~\cite{Schollwock2011}, the MPO representation of two-body terms are easily obtained as follows,
\begin{equation}
\ham_{2b}=\bm{H}_1\mbf{H}_2\cdots\mbf{H}_{N-1}\bm{H}_{N}^T,
\end{equation}
where the local matrix or vector operators are given by,
\begin{align}
\bm{H}_1&=\begin{pmatrix}
0, & \bm{P}_1, & \bm{M}_1, & \bm{Z}_1, & 1 
\end{pmatrix},\\
\mbf{H}_j&=\begin{pmatrix}
1 & & & & \\
\bm{m}_j^T & \mbf{L} & & & \\
\bm{p}_j^T & & \mbf{L} & & \\
\bm{z}_j^T & & & \mbf{L} & \\
0 & \bm{P}_j & \bm{M}_j & \bm{Z}_j & 1 
\end{pmatrix}\hspace{1.5em} (j=2,3,\cdots,N-1),\\
\bm{H}_{N}&=\begin{pmatrix}
1, & \bm{m}_{N}, & \bm{p}_{N}, & \bm{z}_{N}, & 0
\end{pmatrix},
\end{align}
with the lower matrix 
\begin{equation}
\mbf{L}=
\begin{pmatrix}
0 & 0 & 0 & 0 & 0\\
1 & 0 & 0 & 0 & 0\\
0 & 1 & 0 & 0 & 0\\
0 & 0 & 1 & 0 & 0\\
0 & 0 & 0 & 1 & 0\\
\end{pmatrix}.
\end{equation}
Here, we define local vector operators in the dimension of energy for $j=0, 1, 2$ (mod. 3) as follows.
\begin{itemize}
\item $j=0$ (mod. 3)
\begin{align}
\bm{P}_{j}&=\frac{1}{2}
\begin{pmatrix}
J_2^\pri , & J_2^\pri, & J_1^\pri, & 0, & 0
\end{pmatrix}S_{j}^+,\\
\bm{M}_{j}&=\frac{1}{2}
\begin{pmatrix}
J_2^\pri , & J_2^\pri, & J_1^\pri, & 0, & 0
\end{pmatrix}S_{j}^-,\\
\bm{Z}_{j}&=
\begin{pmatrix}
J_2^\pri , & J_2^\pri, & J_1^\pri, & 0, & 0
\end{pmatrix}S_{j}^z.
\end{align}
\item $j=1$ (mod. 3)
\begin{align}
\bm{P}_{j}&=\frac{1}{2}
\begin{pmatrix}
J- \im K/(4\sqrt{3}), & J+ \im K/(4\sqrt{3}) , & J_1^\pri , & J_2^\pri, & J_2^\pri
\end{pmatrix}S_{j}^+,\\
\bm{M}_{j}&=\frac{1}{2}
\begin{pmatrix}
J+ \im K/(4\sqrt{3}), & J- \im K/(4\sqrt{3}) , & J_1^\pri , & J_2^\pri, & J_2^\pri
\end{pmatrix}S_{j}^-,\\
\bm{Z}_{j}&=
\begin{pmatrix}
J, & J, & J_1^\pri , & J_2^\pri, & J_2^\pri
\end{pmatrix}S_{j}^z.
\end{align}
\item $j=2$ (mod. 3)
\begin{align}
\bm{P}_{j}&=\frac{1}{2}
\begin{pmatrix}
J- \im K/(4\sqrt{3}), & J_2^\pri , & J_1^\pri, & J_2^\pri, & 0
\end{pmatrix}S_{j}^+,\\
\bm{M}_{j}&=\frac{1}{2}
\begin{pmatrix}
J+ \im K/(4\sqrt{3}), & J_2^\pri , & J_1^\pri, & J_2^\pri, & 0
\end{pmatrix}S_{j}^-,\\
\bm{Z}_{j}&=
\begin{pmatrix}
J, & J_2^\pri , & J_1^\pri, & J_2^\pri, & 0
\end{pmatrix}S_{j}^z.
\end{align}
\end{itemize}
The dimensionless local vector operators are defined by, 
\begin{align}
&\bm{p}_{j}=
\begin{pmatrix}
 S_{j}^+, & 0, & 0, & 0, & 0
\end{pmatrix},\\
&\bm{m}_{j}=
\begin{pmatrix}
 S_{j}^-, & 0, & 0, & 0, & 0
\end{pmatrix},\\ 
&\bm{z}_{j}=
\begin{pmatrix}
 S_{j}^z, & 0, & 0, & 0, & 0
\end{pmatrix}.
\end{align}

On the other hand, the MPO representation of the three-body terms is neither trivial nor unique.
We use the following form of the MPO representation.
\begin{equation}
\ham_{3b}=\bm{V}_1\mbf{V}_2\cdots\mbf{V}_{N-1}\bm{V}_{N}^T, \label{mpo2}
\end{equation}
where the local vector operators at edge sites are given by,
\begin{align}
&\bm{V}_1=\begin{pmatrix}
\bm{v}_1, & -(2/3)\,KS_1^x, & 0, & 1 
\end{pmatrix},\\
&\bm{V}_{N}=\begin{pmatrix}
\bm{v}_N, & -(2/3)\,KS_N^{-\phi}, & 1, & 0 
\end{pmatrix},
\end{align}
with
\begin{align}
 \bm{v}_{j}=
\begin{pmatrix}
 S_{j}^x, & S_j^y, & S_j^z, & \im/2
\end{pmatrix}.
\end{align}
 The local matrix operators for $j=2,3,\cdots,N-1$ are defined by the following three forms depending on the site index $j$.
 \begin{itemize}
 \item $j=0$ (mod. 3)
 \begin{equation}
   \bm{V}_{j}=\begin{pmatrix}
     \bm{v}_j & -(2/3)\,KS_j^{-\phi} & 1 & 0 \\
     \bm{0}   & 0 & 0 & 1  
   \end{pmatrix}^T.
 \end{equation}
 \item $j=1$ (mod. 3)
 \begin{equation}
   \bm{V}_{j}=\begin{pmatrix}
     \bm{0}   & 0 & 1 & 0 \\
     \bm{v}_j & -(2/3)\,KS_j^{x} & 0 & 1 \\  
   \end{pmatrix}.
 \end{equation}
 \item $j=2$ (mod. 3)
 \begin{equation}
   \bm{V}_{j}=\br{\begin{array}{cc|c}
     -(2/3)\,KS_j^{\phi}\, \mbf{1}_4 & \bm{v}_j^T &  \\
     \bm{v}_j & 0 &  \\ 
     \hline
     & & \mbf{1}_2
   \end{array}},
 \end{equation}
 where $\mbf{1}_4$ ($\mbf{1}_2$) denotes the $4\times 4$ ($2\times 2$) identity matrix.
 \end{itemize}
Note that the MPO representation of a cluster given by a product of three local matrix operators $V_{j}V_{j+1}V_{j+2}$ for $j=1$ (mod. 3) except for $j=1$ and $N-2$ as follows,
\begin{equation}
  V_{j}V_{j+1}V_{j+2}=
  \begin{pmatrix}
    1 & 0\\
    \ham_{3b}^{(j)} & 1
  \end{pmatrix}
\end{equation}
with
\begin{equation}
  \ham_{3b}^{(j)}=-\frac{2K}{3} \BR{S_1^x\, \mrm{Hs}_j(2,3)+S_2^\phi\, \mrm{Hs}_j(3,1)+S_3^{-\phi}\, \mrm{Hs}_j(1,2)}.
\end{equation}
By using this form, we can easily confirm the correspondence of the three-body Hamiltonian \equref{ham3b} and its MPO representation \equref{mpo2}.

\bibliography{refs-a}

\end{document}